\documentclass[10pt]{article}
\usepackage{authblk}
\usepackage[utf8]{inputenc}
\usepackage[T1]{fontenc}
\usepackage{amsmath}
\usepackage{amsfonts}
\usepackage{amssymb}
\usepackage[version=4]{mhchem}
\usepackage{stmaryrd}
\usepackage{hyperref}
\usepackage{hyperref}

\hypersetup{colorlinks=true, linkcolor=blue, filecolor=magenta, urlcolor=cyan,}
\begin{document}
\title{QDistRnd: A GAP package for computing the distance of quantum error-correcting codes}
\author[1,3]{Leonid P. Pryadko}
\author[3]{Vadim A. Shabashov}
\author[2,3]{Valerii K. Kozin}
\affil[1]{Department of Physics \& Astronomy, University of California, Riverside, California, 92521 USA}
\affil[2]{Department of Physics, University of Basel, Klingelbergstrasse 82, CH-4056 Basel, Switzerland}
\affil[3]{The Department of Physics \& Engineering, ITMO University, St. Petersburg, 197101 Russia}

\date{}

\maketitle

Published in JOSS: \href{https://joss.theoj.org/papers/10.21105/joss.04120}{10.21105/joss.04120}

Repository on GitHub \href{https://github.com/QEC-pages/QDistRnd/}{QEC-pages/QDistRnd/}

\section{Summary}
The GAP package QDistRnd implements a probabilistic algorithm for finding the minimum distance of a quantum low-density parity-check code linear over a finite field $\operatorname{GF}(q)$. At each step several codewords are randomly drawn from a distribution biased toward smaller weights. The corresponding weights are used to update the upper bound on the distance, which eventually converges to the minimum distance of the code. While there is no performance guarantee, an empirical convergence criterion is given to estimate the probability that a minimum weight codeword has been found. In addition, a format for storing matrices associated with $q$-ary quantum codes is introduced and implemented via the provided import/export functions. The format, MTXE, is based on the well established MaTrix market eXchange (MTX) Coordinate format developed at NIST, and is designed for full backward compatibility with this format. Thus, MTXE files are readable by any software package which supports MTX.

\section{Statement of need}
Multi-particle quantum correlations can be destroyed rapidly in the presence of errors due to noise, environment, or just random control errors (Nielsen \& Chuang, 2000 \cite{Nielsen}). Quantum error correction (QEC) gives a unique way of controlling such errors and enables, at least theoretically, an arbitrarily long quantum computation when error probability $p$ is below certain threshold, $p_{c}>0$.

QEC requires the use of specially designed quantum error-correcting codes (QECCs). One of the most important parameters of a QECC is the code distance, the minimum weight of a non-trivial logical operator in the code. While for some code families the distance is known or can be related to that of a classical linear error-correcting code, as, e.g., in the case of hypergraph-product and related codes (Tillich \& Zémor, 2009 \cite{Tillich}; Zeng \& Pryadko, 2019, 2020 \cite{Zeng2019,Zeng2020}), in many cases the distance has to be computed directly (Bravyi \& Hastings, 2014 \cite{Bravyi}; Guth \& Lubotzky, 2014 \cite{Guth}; Kovalev \& Pryadko, 2013b \cite{Kovalev2013b}; Panteleev \& Kalachev, 2021b \cite{Panteleev2021b}). Computing the distance is related to the problem of minimum-weight syndrome-based decoding; just like for the classical linear codes (Evseev, 1983 \cite{Evseev}), this problem is NP-hard (note that truly optimal maximum-likelihood decoding for quantum codes requires degeneracy to be taken into account and is a \#P-complete problem (lyer \& Poulin, 2015 \cite{lyer})).

To our knowledge, there is no freely available software for computing the distance of a $q$-ary quantum stabilizer code. A version of the Zimmermann algorithm for finding the distance of linear codes is implemented in Magma (Bosma et al., 1997~\cite{Bosma}), and has been adapted in application to quantum codes, see \href{http://magma.maths.usyd.edu.au/magma/handbook/text/1971#22279}{http://magma.maths.usyd.edu.au/magma/handbook/text/1971\#22279}. Its performance, in particular, in application to practically important (Kovalev \& Pryadko, 2013a \cite{Kovalev2013a}) highly-degenerate quantum codes, also known as quantum LDPC codes, has not been tested by the authors. Several C and C\texttt{++} programs for computing the minimum distance of qubit (binary) Calderbank-Shor-Steane (CSS) codes in various stages of development can also be found at the GitHub respository QEC-pages, owned by one of the authors.

The lack of available software has caused researchers in the field of QECC to either skip the minimum distance calculations altogether (Panteleev \& Kalachev, 2021b \cite{Panteleev2021b}), or develop their own suboptimal algorithms. In particular, Bravyi and Hastings (Bravyi \& Hastings, 2014 \cite{Bravyi}) used an exhaustive search over all non-trivial codewords for calculating the minimum distances.

Note that for some families of QECCs, the distance can be calculated efficiently. In particular, N. P. Breuckmann (2017) \cite{Breuckmann2017} described an algorithm attributed to S. Bravyi for computing the distance of a surface code based on a locally planar graph; for such a code of length $n$ with $k$ logical qubits, the distance can be computed in $\mathcal{O}\left(k n^{2} \log n\right)$ steps. Similarly, a version of the error-impulse method (Declercq \& Fossorier, 2008 \cite{Declercq}; Hu et al., 2004 \cite{Hu}) based on the belief propagation decoding algorithm designed for linear LDPC codes can in principle be used for quantum LDPC codes. We are not aware of any applications of such a technique to QECCs.

We should mention recent theoretical constructions that prove the existence of families of quantum LDPC codes with stabilizer generators of bounded weight and linear (or almost linear) minimum distances (Nikolas P. Breuckmann \& Eberhardt, 2021 \cite{Breuckmann2021}; Hastings et al., 2021 \cite{Hastings2021}; Panteleev \& Kalachev, 2021a, 2022 \cite{Panteleev2021a,Panteleev2022}). Hardly any of the codes from the described families have been explicitly constructed, the reason being that the constructions are expected to produce very long codes. Thus, there is also a need to develop software for calculating minimal distances of quantum codes and optimized specifically for long $\left(n>10^{3}\right)$ and very long $\left(n>10^{5}\right)$ quantum LDPC codes based on qubits.

\section{Functionality of the package}
The distance-finding routines in the package QDistRnd are derived from the code originally written by one of the authors. Implemented algorithm is a variant of the random Information Set (IS) algorithm based on random column permutations and Gauss elimination (Coffey \& Goodman, 1990 \cite{Coffey}; Kruk, 1989 \cite{Kruk}; Leon, 1988 \cite{Leon}). Its eventual convergence for quantum stabilizer codes can be proved based on the existence (Cuéllar et al., 2021 \cite{Cuellar}) of a permutation matrix $P$ such that the reduced row echelon form of the matrix $G^{\prime}=G P$ contains a vector with the weight equal to the distance of the linear code generated by the rows of $G$. Further, a related Covering Set (CS) algorithm has a provable performance (Dumer et al., 2017 \cite{Dumer2017}) for generic (non-LDPC) quantum codes based on random matrices; the corresponding estimate of the number of iterations needed to obtain the distance with probability sufficiently close to 1 also applies for the IS algorithm.

The GAP computer algebra system was chosen because of its excellent support for linear algebra over finite fields. The package QDistRnd gives a reference implementation of the algorithm, with a focus on generality and matrix formats, but not necessarily performance. Nevertheless, the routines are sufficiently fast when dealing with codes of practically important block lengths $n \lesssim 10^{3}$.

The package also contains functions for importing/exporting matrices with elements in a given (finite) Galois field, and a description of a text-based format MTXE based on the well-established MaTrix market eXchange (MTX) Coordinate format developed at NIST (National Institute of Standards and Technology, 2013 \cite{mmformat}). The extension is implemented via structured comments, which guarantees full backward compatibility with the original MTX format. Thus, MTXE files can be read directly by any software package that supports MTX, although some additional processing of matrix elements may be required.

\section{Acknowledgements}
We are grateful to llya Dumer for multiple helpful discussions on the subject. L.P.P. was financially supported in part by the NSF Division of Physics via grants 1820939 and 2112848, and by the Government of the Russian Federation through the ITMO Fellowship and Professorship Program. V.K.K. acknowledges the support from the Georg H. Endress foundation.

{}


\begin{thebibliography}{}
\bibitem{Bosma}
Bosma, W., Cannon, J., \& Playoust, C. (1997). The Magma algebra system. I. The user language. J. Symbolic Comput., 24(3-4), 235-265. \href{https://doi.org/10.1006/jsco.1996.0125}{https://doi.org/10.1006/jsco.1996.0125}

\bibitem{Bravyi}
Bravyi, S., \& Hastings, M. B. (2014). Homological product codes. Proc. Of the 46th ACM Symposium on Theory of Computing (STOC 2014), 273-282. \href{https://arxiv.org/abs/1311.0885}{https://arxiv.org/abs/1311.0885}

\bibitem{Breuckmann2017}
Breuckmann, N. P. (2017). Homological quantum codes beyond the toric code [PhD thesis, RWTH Aachen University]. \href{https://doi.org/10.18154/RWTH-2018-01100}{https://doi.org/10.18154/RWTH-2018-01100}

\bibitem{Breuckmann2021}
Breuckmann, Nikolas P., \& Eberhardt, J. N. (2021). Balanced product quantum codes. IEEE Transactions on Information Theory, 67(10), 6653-6674. \href{https://doi.org/10.1109/TIT.2021.3097347}{https://doi.org/10.1109/TIT.2021.3097347}

\bibitem{Coffey}
Coffey, J. T., \& Goodman, R. M. (1990). The complexity of information set decoding. IEEE Trans. Info. Theory, 36(5), 1031-1037. \href{https://doi.org/10.1109/18.57202}{https://doi.org/10.1109/18.57202}

\bibitem{Cuellar}
Cuéllar, M. P., Gómez-Torrecillas, J., Lobillo, F. J., \& Navarro, G. (2021). Genetic algorithms with permutation-based representation for computing the distance of linear codes. Swarm and Evolutionary Computation, 60, 100797. \href{https://doi.org/10.1016/j.swevo.2020.100797}{https://doi.org/10.1016/j.swevo.2020.100797}

\bibitem{Declercq}
Declercq, D., \& Fossorier, M. (2008). Improved impulse method to evaluate the low weight profile of sparse binary linear codes. Information Theory, 2008. ISIT 2008. IEEE International Symposium on, 1963-1967. \href{https://doi.org/10.1109/ISIT.2008.4595332}{https://doi.org/10.1109/ISIT.2008.4595332}

\bibitem{Dumer2017}
Dumer, I., Kovalev, A. A., \& Pryadko, L. P. (2017). Distance verification for classical and quantum LDPC codes. IEEE Trans. Inf. Th., 63(7), 4675-4686. \href{https://doi.org/10.1109/TIT.2017.2690381}{https://doi.org/10.1109/TIT.2017.2690381} 

\bibitem{Evseev}
Evseev, G. S. (1983). Complexity of decoding for linear codes. Probl. Peredachi Informacii, 19, 3-8. \href{http://mi.mathnet.ru/ppi1159}{http://mi.mathnet.ru/ppi1159}

\bibitem{Guth}
Guth, L., \& Lubotzky, A. (2014). Quantum error correcting codes and 4-dimensional arithmetic hyperbolic manifolds. Journal of Mathematical Physics, 55(8), 082202. \href{https://doi.org/10.1063/1.4891487}{https://doi.org/10.1063/1.4891487} 

\bibitem{Hastings2021}
Hastings, M. B., Haah, J., \& O'Donnell, R. (2021). Fiber bundle codes: Breaking the $N^{1 / 2} \operatorname{poly} \log (N)$ barrier for quantum LDPC codes. STOC 2021: Proceedings of the 53rd Annual ACM SIGACT Symposium on Theory of Computing, 1276-1288. \href{https://doi.org/10.1145/3406325.3451005}{https://doi.org/10.1145/3406325.3451005}

\bibitem{Hu}
Hu, X.-Y., Fossorier, M. P. C., \& Eleftheriou, E. (2004). On the computation of the minimum distance of low-density parity-check codes. Communications, 2004 IEEE International Conference on, 2, 767-771. \href{https://doi.org/10.1109/ICC.2004.1312605}{https://doi.org/10.1109/ICC.2004.1312605}

\bibitem{lyer}
lyer, P., \& Poulin, D. (2015). Hardness of decoding quantum stabilizer codes. IEEE Transactions on Information Theory, 61(9), 5209-5223. \href{https://doi.org/10.1109/TIT.2015.2422294}{https://doi.org/10.1109/TIT.2015.2422294} 

\bibitem{Kovalev2013a}
Kovalev, A. A., \& Pryadko, L. P. (2013a). Fault tolerance of quantum low-density parity check codes with sublinear distance scaling. Phys. Rev. A, 87, 020304(R).  \href{https://doi.org/10.1103/PhysRevA.87.020304}{https://doi.org/10.1103/PhysRevA.87.020304}

\bibitem{Kovalev2013b}
Kovalev, A. A., \& Pryadko, L. P. (2013b). Quantum Kronecker sum-product low-density parity-check codes with finite rate. Phys. Rev. A, 88, 012311. \href{https://doi.org/10.1103/PhysRevA.88.012311}{https://doi.org/10.1103/PhysRevA.88.012311}

\bibitem{Kruk}
Kruk, E. A. (1989). Decoding complexity bound for linear block codes. Probl. Peredachi Inf., 25(3), 103-107. \href{http://mi.mathnet.ru/eng/ppi665}{http://mi.mathnet.ru/eng/ppi665}

\bibitem{Leon}
Leon, J. S. (1988). A probabilistic algorithm for computing minimum weights of large error-correcting codes. IEEE Trans. Info. Theory, 34(5), 1354-1359. \href{https://doi.org/10.1109/18.21270}{https://doi.org/10.1109/18.21270}

\bibitem{mmformat}
National Institute of Standards and Technology. (2013). Matrix market exchange formats. online. \href{https://math.nist.gov/MatrixMarket/formats.html}{https://math.nist.gov/MatrixMarket/formats.html}

\bibitem{Nielsen}
Nielsen, M. A., \& Chuang, I. L. (2000). Quantum computation and quantum infomation. Cambridge Unive. Press.

\bibitem{Panteleev2021a}
Panteleev, P., \& Kalachev, G. (2021a). Asymptotically good quantum and locally testable classical LDPC codes. \href{http://arxiv.org/abs/2111.03654}{http://arxiv.org/abs/2111.03654}

\bibitem{Panteleev2021b}
Panteleev, P., \& Kalachev, G. (2021b). Degenerate quantum LDPC codes with good finite length performance. Quantum, 5, 585. \href{https://doi.org/10.22331/q-2021-11-22-585}{https://doi.org/10.22331/q-2021-11-22-585}

\bibitem{Panteleev2022}
Panteleev, P., \& Kalachev, G. (2022). Quantum LDPC codes with almost linear minimum distance. IEEE Transactions on Information Theory, 68(1), 213-229. \href{https://doi.org/10.1109/TIT.2021.3119384}{https://doi.org/10.1109/TIT.2021.3119384}

\bibitem{Tillich}
Tillich, J.-P., \& Zémor, G. (2009). Quantum LDPC codes with positive rate and minimum distance proportional to $\sqrt{n}$. Proc. IEEE Int. Symp. Inf. Theory (ISIT), 799-803. \href{https://doi.org/10.1109/ISIT.2009.5205648}{https://doi.org/10.1109/ISIT.2009.5205648}

\bibitem{Zeng2019}
Zeng, W., \& Pryadko, L. P. (2019). Higher-dimensional quantum hypergraph-product codes with finite rates. Phys. Rev. Lett., 122, 230501. \href{https://doi.org/10.1103/PhysRevLett.122.230501}{https://doi.org/10.1103/PhysRevLett.122.230501}

\bibitem{Zeng2020}
Zeng, W., \& Pryadko, L. P. (2020). Minimal distances for certain quantum product codes and tensor products of chain complexes. Phys. Rev. A, 102, 062402. \href{https://doi.org/10.1103/PhysRevA.102.062402}{https://doi.org/10.1103/PhysRevA.102.062402}
\end{thebibliography}
\end{document}